\documentclass[letterpaper,prb,superscriptaddress,twocolumn,showpacs]{revtex4}
\usepackage{graphicx}
\usepackage{bm}
\usepackage{amsmath}
\usepackage{amsbsy}
\usepackage{amssymb}
\usepackage{array}
\usepackage{wasysym}

\begin{document}
\title{Ruderman-Kittel-Kasuya-Yosida interactions on a bipartite lattice}
\author{J. E. Bunder}
\affiliation{Nanomechanics Group, School of Mathematics and Applied Statistics, University of Wollongong, Wollongong, NSW 2522, Australia}
\author{Hsiu-Hau Lin}
\affiliation{Department of Physics, National Tsing-Hua University, Hsinchu 300, Taiwan}
\affiliation{Physics Division, National Center for Theoretical Sciences, Hsinchu 300, Taiwan}

\begin{abstract}
Carrier-mediated exchange coupling, known as Ruderman-Kittel-Kasuya-Yosida (RKKY) interaction, plays a fundamental role in itinerant ferromagnetism and has great application potentials in spintronics. A recent theorem based on the imaginary-time method shows that the oscillatory RKKY interaction becomes commensurate on bipartite lattice and predicts that the effective exchange coupling is always ferromagnetic for the same sublattice but antiferromagnetic for opposite sublattices. We revisit this important problem by real- and imaginary-time methods and find the theorem misses important contributions from zero modes. To illustrate the importance of zero modes, we study the spin susceptibility in graphene nanoribbons numerically. The effective exchange coupling is largest on the edges but does not follow the predictions from the theorem.
\end{abstract}


\maketitle


Exchange coupling between localized magnetic moments leads to various magnetic phases that are important for spintronics applications.\cite{Wolf01,Zutic04,MacDonald05} Unlike the direct exchange coupling, the Ruderman-Kittel-Kasuya-Yosida (RKKY) interaction is mediated by itinerant carriers in the host material and is sensitive to the low-energy properties near the Fermi surface.\cite{Hindmarch03} On a bipartite lattice, the Fermi surface is naturally nested. Thus, it is interesting to explore the interplay between the lattice structure and the oscillations in the RKKY interaction. A recent theorem\cite{Saremi07} predicts that the oscillatory RKKY interaction is commensurate on a bipartite lattice, so that the effective exchange coupling between two localized moments is always ferromagnetic when they are on the same sublattice, but antiferromagnetic for opposite sublattices.

Graphene-based materials\cite{Novoselov04,Novoselov05,Zhang05,Zhou06} are ideal candidates for exploring these peculiar features. Firstly, unlike most conventional metals, the Fermi surface of graphene shrinks to two Dirac points with relativistic dispersion,\cite{Geim07} and the electronic structure is well described by the tight-binding model on the honeycomb lattice. In addition, since the successful fabrication of planar graphene it has become more than just an academic curiosity, as there is great application potential for electronic transport at the nanoscale\cite{Berger06,Miao07}. Of particular interest are studies and proposals which demonstrate applications for graphene nanoribbons in the field of spintronics.\cite{Rycerz07,Trauzettel07} For example, graphene nanoribbons with zigzag edges support localized states at the Fermi energy and cause the accumulation of spin and charge polarization near the edges.\cite{Fujita96,Nakada96,Wakabayashi98,Hikihara03,Lee05,Yazyev08}

\begin{figure}
\centering
\includegraphics[width=7.5cm]{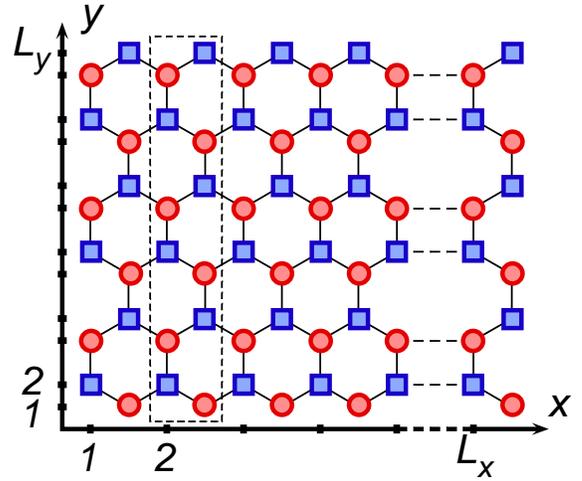}
\caption{(Color Online) A zigzag graphene nanoribbon with width $L_y=12$,  length $L_x$ and sublattices represented by squares and circles. Each site is defined by integral values of $x$ and $y$, e.g., the rectangle contains all sites with $x=2$.}
\label{fig:lattice}
\end{figure}

Here we aim to better understand the carrier-mediated exchange coupling on bipartite lattice and use a graphene nanoribbon as a demonstrating example. We calculate the carrier-mediated exchange coupling using the real-time formalism and find the results do not agree with the theorem in Ref. \onlinecite{Saremi07} obtained from the imaginary-time approach. The discrepancy between these two methods is analyzed in detail. The key lies in the zero modes of the system which require extra care to be taken with the analytic continuation connecting the imaginary-time Green's function to the real one. To illustrate the discrepancy between the two approaches explicitly, we numerically compute the RKKY interaction on the zigzag graphene nanoribbon shown in Fig. \ref{fig:lattice}. Our results clearly show that the effective exchange coupling is not always ferromagnetic for two moments on the same sublattice, and neither is it always antiferromagnetic on opposite sublattices.


We start with the general hopping Hamiltonian on a bipartite lattice,
\begin{equation}
H = \sum_{i,j} t_{ij} c^\dag_i c_j,\label{eq:H}
\end{equation}
where $t_{ji}=t^*_{ij}$ to ensure hermicity, and the spin indices are suppressed. For simplicity, we will only consider the real hopping amplitude $t_{ij}$. The bipartite condition requires that $t_{ij}$ is non-vanishing only for $i, j$ on different sublattices $A, B$. The chemical potential $\mu$ is set to zero so that the system is at half-filling. Undoped graphene is at half-filling, and can be described by an effective tight-binding model on a bipartite lattice, such as that given in Eq. (\ref{eq:H}).

The carrier-mediated exchange coupling is proportional to the static spin susceptibility,
\begin{equation}
J_{ij} = -J^2 \chi^R_{ij} (\omega=0),
\end{equation}
where $J$ is the coupling between the impurity spins and the itinerant carriers on the lattice. The static spin susceptibility can be obtained from the Fourier transform of the retarded spin susceptibility,
\begin{equation}
\chi^R_{ij}(t) = \frac{i}{2} \Theta(t) \Big\langle S^-_i(t) S^+_j(0)
-S^+_j(0) S^-_i(t) \Big\rangle.
\end{equation}
We can compute the correlation function directly in real time, or we can use the imaginary-time method by analytic continuation.

Since the hopping amplitude is spin independent, the correlation functions do not depend on the spin orientations. Furthermore, the bipartite condition gives rise to particle-hole symmetry and leads to the useful relation, $\langle c_{i}(t) c^\dag_{j}(0) \rangle = \epsilon_i \epsilon_j \langle c^\dag_{i}(t) c_{j}(0) \rangle$, where $\epsilon_i =1$ for $i \in A$ and $\epsilon_i=-1$ for $i \in B$. After some algebra, the spin correlation function takes the simple form,
\begin{equation}
\langle S^-_i(t) S^+_j(0) \rangle = \epsilon_i \epsilon_j G^2_{ij}(t),
\end{equation}
where $G_{ij}(t) \equiv \langle c^\dag_i(t) c_j(0) \rangle$ is the single-particle correlation function. After a Fourier transform and setting $\omega=0$, the carrier-mediated exchange coupling can be expressed as an integral,
\begin{equation}
J_{ij} = J^2 \epsilon_i \epsilon_j \int_0^\infty dt\:
{\rm Im} [G^2_{ij}(t)],
\end{equation}
where the relation $G_{ji}(-t) = [G_{ij}(t)]^*$ has been used.

The single-particle correlation function can be expanded,
\begin{equation}
G_{ij}(t) = \langle c^\dag_i(t) c_j(0) \rangle =
\sum_n \phi^*_n(i) \phi_n(j) e^{i \xi_n t} n_F(\xi_n),
\end{equation}
where $\phi_n(i)$ is the eigenfunction with energy $\xi_n$ and $n_F$ is the Fermi distribution function. Since the hopping amplitude is real, for a given $n$ the eigenfunctions $\phi_n(i)$ can be chosen to be wholly real or wholly imaginary for all $i$. On evaluating the integral one introduces the convergence factor $e^{-\eta t}$ and
\begin{eqnarray}
J_{ij} \hspace{-1mm} &=& \hspace{-1mm} J^2 \epsilon_i \epsilon_j \sum_{n,m} W_{nm}(i,j) n_F(\xi_n) n_F(\xi_m)
\frac{\xi_n+\xi_m}{(\xi_n+\xi_m)^2+\eta^2}
\nonumber\\
&=& J^2 \epsilon_i \epsilon_j \sum_{\xi_n+\xi_m\neq 0} W_{nm}(i,j)\:
\frac{n_F(\xi_n) n_F(\xi_m)}{\xi_n+\xi_m},\label{eq:J}
\end{eqnarray}
where the product of the eigenfunctions is $W_{nm}(i,j) = \phi^*_n(i) \phi_n(j) \phi^*_m(i) \phi_m(j)$, which is always real. Note that the zero modes with $\xi_n+\xi_m=0$ do not contribute to the RKKY interaction and that the expression in the second line follows in the $\eta \to 0$ limit.


Now we turn to the imaginary-time approach. We choose $\tau >0$ so that the time ordering is straightforward. The spin susceptibility in the imaginary-time formalism is
\begin{equation}
\chi_{ij}(\tau) = \frac{1}{2}
\bigg\langle T_\tau S^-_i(\tau) S^+_j(0) \bigg\rangle
= \frac12 \epsilon_i \epsilon_j {\cal G}_{ij}^2(\tau),
\end{equation}
where the imaginary-time Green's function is ${\cal G}_{ij}(\tau) = \langle c_i^\dag(\tau) c_j(0) \rangle$ for $\tau >0$. In the Fourier space, the spin susceptibility is
\begin{equation}
\chi_{ij}(i\Omega_n) = \frac12 \epsilon_i \epsilon_j
\int_0^\beta d\tau\: e^{i \Omega_n \tau}
{\cal G}_{ij}^2(\tau).
\end{equation}
If the analytic continuation $i\Omega_n \to \omega + i \eta$, is performed as an initial step then the RKKY interaction obtained from the imaginary-time method is
\begin{eqnarray}
{\cal J}_{ij} = - \frac12 J^2 \epsilon_i \epsilon_j \int_0^\beta d\tau \:
 e^{i\eta \tau} {\cal G}^2_{ij} (\tau).\label{eq:curlyJ1}
\end{eqnarray}
Now, if $\eta\to 0$ the RKKY interaction becomes
\begin{equation}
\mathcal{J}_{ij} = - \frac12 J^2 \epsilon_i \epsilon_j \int_0^\beta d\tau {\cal G}^2_{ij} (\tau).
\label{eq:curlyJ2}
\end{equation}
As the above integral is positive definite, the sign of ${\cal J}_{ij}$ only depends on the product $\epsilon_i \epsilon_j$. If both spins are on the same sublattice, $\epsilon_i \epsilon_j =1$ and the coupling is apparently ferromagnetic. If the spins are on opposite sublattices, $\epsilon_i \epsilon_j=-1$ and the coupling is  antiferromagnetic.

The commensurate feature described by ${\cal J}_{ij}$ may seem reasonable. For example, the tight-binding model on the square lattice at half filling is bipartite with a nesting vector $\bm{Q} = (\pi,\pi)$. The particle-hole excitations near the Fermi surface carry the same momentum and produce an oscillatory factor $\cos \bm{Q} \cdot \bm{r} = (-1)^{x+y}$ in the RKKY interaction. It is clear that the oscillation is commensurate with the underlying lattice, as predicted by the theorem in Ref. \onlinecite{Saremi07}.
However, the analytic continuation $i \Omega_n \to \omega + i\eta$ is established in the Lehmann decomposition, where the integration over imaginary time must be carried out first and the Wick rotation is the last step  of the calculation. When the analytic continuation is performed correctly, $\chi^R_{ij}(z)=\chi_{ij}(z)$ where the frequency $z$ is a complex number in the upper half of the plane, which does not include the real axis, and one can calculate the RKKY interaction using either the real or imaginary time method. If the analytic continuation is performed before the integration, as is done in Eq. (\ref{eq:curlyJ1}) and Eq. (\ref{eq:curlyJ2}), the solution may be incorrect.

Let us elaborate on the results obtained from Eq. (\ref{eq:curlyJ1}). The Greens' function takes the form,
\begin{equation}
{\cal G}_{ij}(\tau) = \sum_n \phi^*_n(i) \phi_n(j) n_F(\xi_n) e^{\tau \xi_n},
\end{equation}
and after integrating over the imaginary time, the exchange coupling is
\begin{eqnarray}
\mathcal{J}_{ij} = \frac{J^2\epsilon_i \epsilon_j}{2}\sum_{n,m}
W_{nm}(i,j) n_F(\xi_n) n_F(\xi_m)\nonumber\\
\times \frac{1-e^{\beta(\xi_n+\xi_m+i \eta)}}{\xi_n+\xi_m+i\eta}.
\label{eq:Jim}
\end{eqnarray}
The sum can be separated into two parts, the first contains contributions from zero modes with $\xi_n + \xi_m = 0$, and the second contains all other terms with $\xi_n + \xi_m \neq 0$. The first part can be shown to equal
\begin{equation}
{\cal J}^0_{ij} = -\frac{J^2 \epsilon_i \epsilon_j }{2}\sum_{\xi_n+\xi_m = 0}
W_{nm}(i,j) \beta n_F(\xi_n) n_F(\xi_m).
\label{eq:zero}
\end{equation}
As for the second term, the particle-hole symmetry ensures that each energy $\xi_n$ is paired with an equal energy of opposite sign $\xi_{\bar{n}}=-\xi_n$, and that their eigenfunctions satisfy $\phi_{\bar{n}}(i)=\epsilon_i\phi_n(i)$. In addition, particle-hole symmetry provides the useful relation $n_F(\xi_{n}) e^{\beta \xi_{n}} = n_F(\xi_{\bar n})$. Therefore,
\begin{eqnarray}
-\sum_{\xi_n+\xi_m \neq 0}
W_{nm}(i,j)
n_F(\xi_n) n_F(\xi_m) \frac{e^{\beta(\xi_n+\xi_m)}}{\xi_n+\xi_m+i\eta}
\nonumber\\
=  \sum_{\xi_n+\xi_m \neq 0} W_{\bar{n} \bar{m}}(i,j) n_F(\xi_{\bar n}) n_F(\xi_{\bar m})
\frac{1}{\xi_{\bar n}+\xi_{\bar m}-i\eta},
\end{eqnarray}
and it can be seen, after exchanging the dummy variables $(\bar{n},\bar{m})$ with $(n,m)$, that the $\xi_m+\xi_n\neq 0$ part of Eq. (\ref{eq:curlyJ1}) is equivalent to Eq. (\ref{eq:J}).
Therefore we have shown $\mathcal{J}_{ij}={\cal J}^0_{ij}+J_{ij}$.

The discrepancy between $J_{ij}$ and ${\cal J}_{ij}$ always exists at finite temperature and, as we will show in the graphene nanoribbon example, there are cases where zero temperature also leads to a discrepancy. We have performed numerical calculations (not shown here) on simple bipartite lattices and verified that the RKKY interaction $J_{ij}$ need not follow the sign rule obeyed by $\mathcal{J}_{ij}$.
However, there are special limits where the correction ${\cal J}^0_{ij}$ vanishes. If all single-particle states are gapped ($\xi_m = -\xi_n \neq 0$) and the temperature approaches zero,
\begin{equation}
\beta n_F(\xi_n) n_F(-\xi_n) \sim \beta e^{-\beta \Delta} \to 0.\label{eq:zeroT}
\end{equation}
Therefore, the difference between $J_{ij}$ and ${\cal J}_{ij}$ vanishes for a gapped system at zero temperature.

In summary, while $\chi^R_{ij}(\omega)$ and $\chi_{ij}(i\Omega_n\to \omega+ i\eta)$ are identical and either can be used to evaluate the RKKY interactions, one must be careful with the analytic continuation when using the imaginary time formulation. The analytic continuation must be taken after the imaginary time integration, and not before, as these two operations do not in general commute. If done correctly, the RKKY interaction will not have any contributions from the zero modes of the system, $\xi_n+\xi_m=0$. However, taking the analytic continuation prior to evaluating the imaginary time integral will results in a zero-mode term $\mathcal{J}^0_{ij}$.

While $\mathcal{J}^0_{ij}$ will always be non-zero at finite temperatures, for systems with a regular density of states near the Fermi surface, the contributions from the zero modes scale to zero in the thermodynamic limit. One can understand this by considering how the terms in Eq. (\ref{eq:zero}) scale with volume $V$. The eigenfunction term scales as $W_{nm} \sim (1/V)^2$. On eliminating the constrained variable $m$, the summation over $n$ scales as $V \int d^d k$ in the thermodynamic limit. Thus, the correction term $\mathcal{ J}^0_{ij}$ scales as $(1/V)^2 \cdot V = 1/V$ and vanishes in the thermodynamic limit. However, there are a number of systems which do not have a regular density of states in the thermodynamic limit. Graphene nanoribbons, for example, have a singular density of states at the Fermi surface and therefore $\mathcal{J}^0_{ij}$ is never zero.


To compute the RKKY interaction for the zigzag graphene nanoribbon, shown in Fig. \ref{fig:lattice}, we first need to obtain all eigenstates. Making use of the translational invariance along the ribbon direction, we take the partial Fourier transform of $x$ to $k_x$. The operators which diagonalize the Hamiltonian are defined by $\psi(\wp)=\sum_y\phi^*(\wp,y)c(k_x,y)$ where $\wp=(k_x,p)$. The solution for $|k_x|<k_0=2\cos^{-1}[L_y/2(L_y+2)]$) is
\begin{eqnarray}
\phi(\wp,y) &=& \pm\sin[py+\varphi(\wp)],\qquad \mbox{for odd $y$},
\nonumber\\
\phi(\wp,y) &=& \sin[py], \hspace{2.2cm} \mbox{for even $y$},
\label{eq:phi}
\end{eqnarray}
where $\varphi(\wp)=\tan^{-1}\left[\tan (p)(t'-t)/(t'+t)\right]$ with range $-\pi/2\leq\varphi\leq\pi/2$, and $t'=2t\cos(k_x/2)$. As $L_y$ is finite we have the constraint $(L_y+1)p+\varphi(\wp)=m\pi$
with $m=1,2,\ldots L_y/2$. The energy is
$E(\wp)=\mp\sqrt{t'^2+t^2+2t't\cos(2p)}$.
Therefore we have $L_y/2$ positive bands and $L_y/2$ negative bands each associated with one of the $L_y/2$ values of $m$. For $|k_x|>k_0$ the solutions for $m=1,2,\ldots (L_y-2)/2$ do not change but $m=L_y/2$ now describes localized edge states with $p=\pi/2+i\gamma$. Substituting this imaginary $p$ into the $|k_x|<k_0$ solution with $m=L_y/2$ gives
\begin{eqnarray}
\phi(\wp,y)& = &\pm e^{i\pi y/2} \sinh[\gamma(L_y+1-y)], \mbox{for odd $y$},
\nonumber\\
\phi(\wp,y)& = &e^{i\pi (y-1)/2} \sinh[\gamma y], \hspace{1.2cm} \mbox{for even $y$},\label{eq:phigamma}
\end{eqnarray}
with $(L_y+1)\gamma=\tanh^{-1}\left[-\tanh(\gamma)(t'+t)/(t'-t) \right]$ and the dispersion for the edge state is $E(\wp)=\mp\sqrt{t'^2+t^2-2t't\cosh(2\gamma)}$. Note that the eigenfunctions of the edge states rapidly decay into the bulk.

\begin{figure}
\centering
\includegraphics[width=8cm]{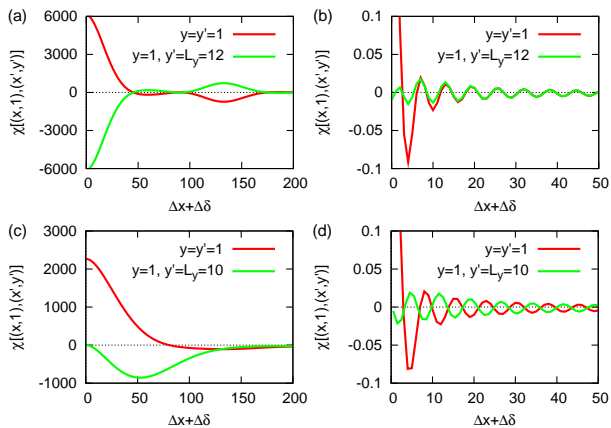}
\caption{(Color Online) The static spin susceptibility $\chi[(x,1),(x',y')]$ for a zigzag graphene nanoribbon at zero temperature. Since it is dominated by the edge response, we choose $y'=1$ and $y'=L_y$ and plot the susceptibility versus the horizontal distance $\Delta x + \Delta \delta$ for a: symmetric  nanoribbon $L_y=12$ (a) at half filling $\mu=0$, (b) at slight doping $\mu=0.01$; asymmetric nanoribbon $L_y=10$ (a) at half filling $\mu=0$, (d) at slight doping $\mu=0.01$.}
\label{fig:susceptibility}
\end{figure}

The translational invariance of the static spin susceptibility $\chi(x,y;x',y')$ implies that it is dependent on the horizontal distance $|x-x'|$. However, a lattice point's $x$ value may not represent its true position and so we introduce the offset $\delta$ where $\delta=1/2$ for $y=1,0\,\mathrm{mod}\,4$ and $\delta=0$ for $y=2,3\,\mathrm{mod}\,4$. Thus, the spin susceptibility depends on $\Delta x+ \Delta \delta = (x-x')+(\delta - \delta')$. The numerical results at half-filling ($\mu=0$) and with slight doping ($\mu=0.01$) for the symmetric ($L_y=12$) and asymmetric ($L_y=10$) lattices are summarized in Fig. \ref{fig:susceptibility}.

Since the spin susceptibility is largest on the edges, we will focus on these lattice sites. When $L_y/2$ is even, the upper and the lower edges are symmetric and the spin susceptibilities for $y=1$ with $y'=1$ and $y'=L_y$ are the same except for the sign. A perfect reflection symmetry is clearly seen in Fig. \ref{fig:susceptibility}(a). Note that for $y=y'=1$ the spins are on the same sublattice (labeled as red circles in Fig. \ref{fig:lattice}), but the RKKY interaction $J_{ij} = - J^2 \chi^R_{ij}$ is not always ferromagnetic, as shown in Fig. \ref{fig:susceptibility}(a). Similarly, for $y=1$ and $y'=L_y$ the spins are on opposite sublattices, but the RKKY interaction is not always antiferromagnetic. In fact, both curves show non-trivial oscillations which pass through zero. This provides clear evidence that the theorem given in Ref. \onlinecite{Saremi07} is incorrect. With only slight doping ($\mu=0.01$), we lose the particle-hole symmetry and conventional RKKY oscillations appear, as shown in Fig. \ref{fig:susceptibility}(b).

For the asymmetric nanoribbon with $L_y/2$ an odd integer, the reflection symmetry between the spin susceptibilities for $y=1$ with $y'=1$ and $y'=L_y$ is lost, as clearly shown in Fig. \ref{fig:susceptibility}(c). However, the RKKY interaction for spins on the same sublattice ($y=y'=1$) is still not consistently ferromagnetic. Upon slight doping ($\mu=0.01$), conventional RKKY oscillations again appear, as shown in Fig. \ref{fig:susceptibility}(d). In the asymptotic regime, the spin susceptibilities for $y'=1$ and $y'=L_y$ are mirror images of each other due to a half lattice shift between the upper and the lower edges of the asymmetric nanoribbon.

In conclusion, we show that the RKKY interaction on a bipartite lattice is not commensurate with the lattice, although it has been predicted to be so. The discrepancy arises when the analytic continuation is performed incorrectly, which leads to errors which can be attributed to the zero modes in the system. The zigzag graphene nanoribbon is studied numerically as an illustrating example which highlights the discrepancy.

We acknowledge the support of the National Science Council of Taiwan through grant No. NSC-97-2112-M-007-022-MY3 and also partial support from the National Center for Theoretical Sciences in Taiwan.

\end{document}